\documentclass[%
 aip,
 amsmath,amssymb,
 reprint,%
]{revtex4-1}

\usepackage{graphicx}
\usepackage{dcolumn}
\usepackage{bm}

\usepackage[utf8]{inputenc}
\usepackage[T1]{fontenc}
\usepackage{amsmath}

\usepackage{xcolor}
\usepackage{amssymb}
\usepackage{mathptmx}
\usepackage{braket}
\usepackage{etoolbox}

\makeatletter
\def\@email#1#2{%
 \endgroup
 \patchcmd{\titleblock@produce}
  {\frontmatter@RRAPformat}
  {\frontmatter@RRAPformat{\produce@RRAP{*#1\href{mailto:#2}{#2}}}\frontmatter@RRAPformat}
  {}{}
}%
\makeatother
\begin{document}

\preprint{AIP/123-QED}

\title[Geometric Control of Electronic Decoherence in a Chiral Spin-Orbit Coupled Helix]{Geometric Control of Electronic Decoherence in a Chiral Spin-Orbit Coupled Helix}
\author{Neel Moudgal}
    \affiliation{Department of Biophysics, University of Michigan, Ann Arbor, Michigan 48109, USA}
  \affiliation{Department of Chemical Engineering, University of Michigan, Ann Arbor, Michigan 48109, USA}
  \affiliation{Department of Chemistry, University of Michigan, Ann Arbor, Michigan 48109, USA}
\author{Nicholas A. Kotov}%
   \affiliation{Department of Chemical Engineering, University of Michigan, Ann Arbor, Michigan 48109, USA}
      \email{kotov@umich.edu}
\author{Eitan Geva}
\affiliation{Department of Chemistry, University of Michigan, Ann Arbor, Michigan 48109, USA}
\email{eitan@umich.edu}

\date{\today}

\begin{abstract}
While 
chirality is known to impact chemical and optical properties of molecular systems, as well as their biological function, 
far less is known about its impact on quantum coherence. 
By considering a chiral helix
as a paradigmatic model, we show that within the Born-Markov regime, changing the
degree of helical twist can be used to modulate the electronic decoherence rate. 
Moderate to strong spin-orbit coupling between next-nearest-neighbor sites along the helix are shown to give rise to a strong dependence of the electronic decoherence rate on the degree of chirality. 
The effect of chirality on electronic decoherence rates is also seen to be strongly impacted by the spatial and temporal correlations between the local environmental fluctuations to which the local excitations along the helix are subject.
Our results show that chirality
can serve as a tunable design parameter for quantum engineering and 
highlight the importance role environmental correlations can play in determining how chirality affects electronic decoherence rates. 
\end{abstract}

\maketitle
\section{Introduction}
Chirality, the mirror asymmetry of molecules, is an important aspect of molecular structure that dictates the behavior of chemical and biological systems.\cite{sakhno_chirality_2025} Indeed, biological systems are well known to be homochiral with living systems containing a preponderance of L-amino acids and D-sugars.\cite{laurent_emergence_2021, ozturk_lifes_2025} As a result, biological macromolecules have characteristic chiral geometries that dictate their ability to recognize and interact with chiral substrates and other macromolecules.\cite{moudgal_multiscale_2025}

In addition to organic and biomolecular structures, inorganic structures can also be chiral and exhibit chirality-dependent properties. For example, engineered nanostructures  exhibit unique properties, such as strong optical responses to circularly polarized light, as a result of structural chirality.\cite{kotov_chiral_2022, kotov_inorganic_2010, lu_emerging_2021} An important finding of studies involving chiral nanostructures is that chirality is not simply a binary property: it is a property that manifests differently across various length scales.\cite{moudgal_multiscale_2025} In particular, a single nanostructure (including biomolecules) may manifest both left-and right-handed features at different scales which translates to varied properties across scales (e.g. optical response, interactions). \cite{jiang_emergence_2020, shen_chiral_2021}

Work on the mathematical theory of chirality and chirality measures has demonstrated that chirality itself is a property that can be quantified in a variety of systems.\cite{osipov_new_1995, buda_hausdorff_1992, tuvi-arad_csm_2024, cha_graph-theoretical_2024} A number of chirality measures have been proposed, each with a different set of properties that make them suited for different classes of systems. \cite{millar_osipovpickupdunmur_2005, ni_seed-mediated_2024} However, all chirality measures share a notable common feature: they represent the degree of structural asymmetry as a continuous variable rather than a discrete state. This continuous nature of chirality has also been experimentally realized in studies of chiral nanostructures across a variety of length scales. For example the twist of small chiral organic molecules (e.g. helicenes and $\pi$-extended twistacenes) can be varied by systematically altering synthetic functionalization.\cite{bedi_helically_2018, suarez-pantiga_-fjord_2021} Highly tunable geometric chirality has been elegantly demonstrated in self-assembled nanostystems in which twist can be modulated by varying synthesis conditions. \cite{kumar_photonically_2023} Crucially, across these diverse molecular architectures, the observed physical and quantum characteristics do not simply toggle between discrete states. Instead, they exhibit well-defined chirality-property relations, where the system's unique optical and electronic properties are seen to continuously vary with the underlying chiral morphology.

A large body of work has demonstrated that structural chirality can play a key role in quantum mechanical phenomena. The chirality-induced spin selectivity (CISS) effect, by which electron transport through a helical structure results in a spin-polarized current, has been verified in a variety of systems, including organic systems, biomolecular systems, and donor-acceptor molecular systems.\cite{ray_asymmetric_1999, gohler_spin_2011, naaman_chiral_2022, bloom_chiral_2024, eckvahl_direct_2023} Although a unified mechanistic picture explaining the physical origin of this effect remains elusive, what is clear is that structural chirality can impact quantum mechanical observables like electron spin.\cite{evers_theory_2022, nuomin_theories_2026} 

While the interplay between chiral geometry and quantum mechanical observables has been extensively investigated in non-equilibrium transport phenomena (such as CISS), its role in the dynamics of open quantum systems remains largely unexplored. If helical geometry dictates the internal spin-orbit coupling (SOC), it must inherently mediate how coherent electronic superpositions decay when coupled to an external environment (e.g. a phonon bath). We therefore focus attention here on the means by which the chiral geometry (i.e. degree of twist) of a helix dictates rate of electronic decoherence. 

To systematically investigate how chiral geometry influences electronic decoherence rates, a helical lattice serves as a paradigmatic model. By modulating the structural twist of the helix—defined by the azimuthal step between adjacent sites—while preserving local nearest-neighbor bond lengths, one can smoothly and continuously tune the degree of chiral asymmetry. This makes the helix an ideal system for examining how scale-resolved geometric changes manifest in physical observables. Crucially, this continuous geometric tuning directly governs the local quantum mechanical interactions within the structure. In a tight-binding framework, next-nearest-neighbor SOC is highly sensitive to the three-dimensional arrangement of the lattice, scaling with the cross-product of adjacent bond vectors.\cite{fransson_vibrational_2020} Consequently, altering the helical twist provides a direct, geometric mechanism to continuously modulate the effective SOC experienced by an electron.

Furthermore, evaluating this geometric control of electronic decoherence necessitates moving beyond standard collective bath approximations often used in the study of electron transport \cite{fransson_vibrational_2020, upadhyay_weak_2026} and instead considering local baths acting on each site of the helix. From these local baths, one may further examine how bath correlations—both across spatial sites and distinct electronic channels—affects the relationship between chiral geometry and decoherence. 

In this work, we present a generalized local-coupling model to demonstrate that helical geometry fundamentally controls the rate of electronic decoherence. By computing Redfield tensor coefficients to extract decoherence rates of a prepared electronic superposition, we show that the nature of this control is modulated by a number of factors including SOC strength and bath correlations. Although the results presented here are for an idealized helical lattice, we anticipate that the underlying principles will translate broadly as a design principle for physically realizable small quantum systems. Ultimately, this framework provides a first investigation of the so-far unexplored bridge between continuous structural chirality and the decoherence rates.

The remainder of this paper is organized as follows. In Sec. II, we introduce the chiral tight-binding model, describe the system-bath coupling and bath correlation structure, and outline the Redfield theory used to calculate electronic decoherence rates. In Sec. III, we examine the dependence of decoherence on helical geometry, spin–orbit coupling strength, and bath correlations. We then investigate the physical origins of these trends through the contributions of pure dephasing and population relaxation and assess their robustness across different helix lengths. Finally, in Sec. IV, we summarize our principal findings and their implications for geometric control of decoherence in quantum systems.

\section{Theory and Methods}

\begin{figure}
    \centering
    \includegraphics[width=1.0\linewidth]{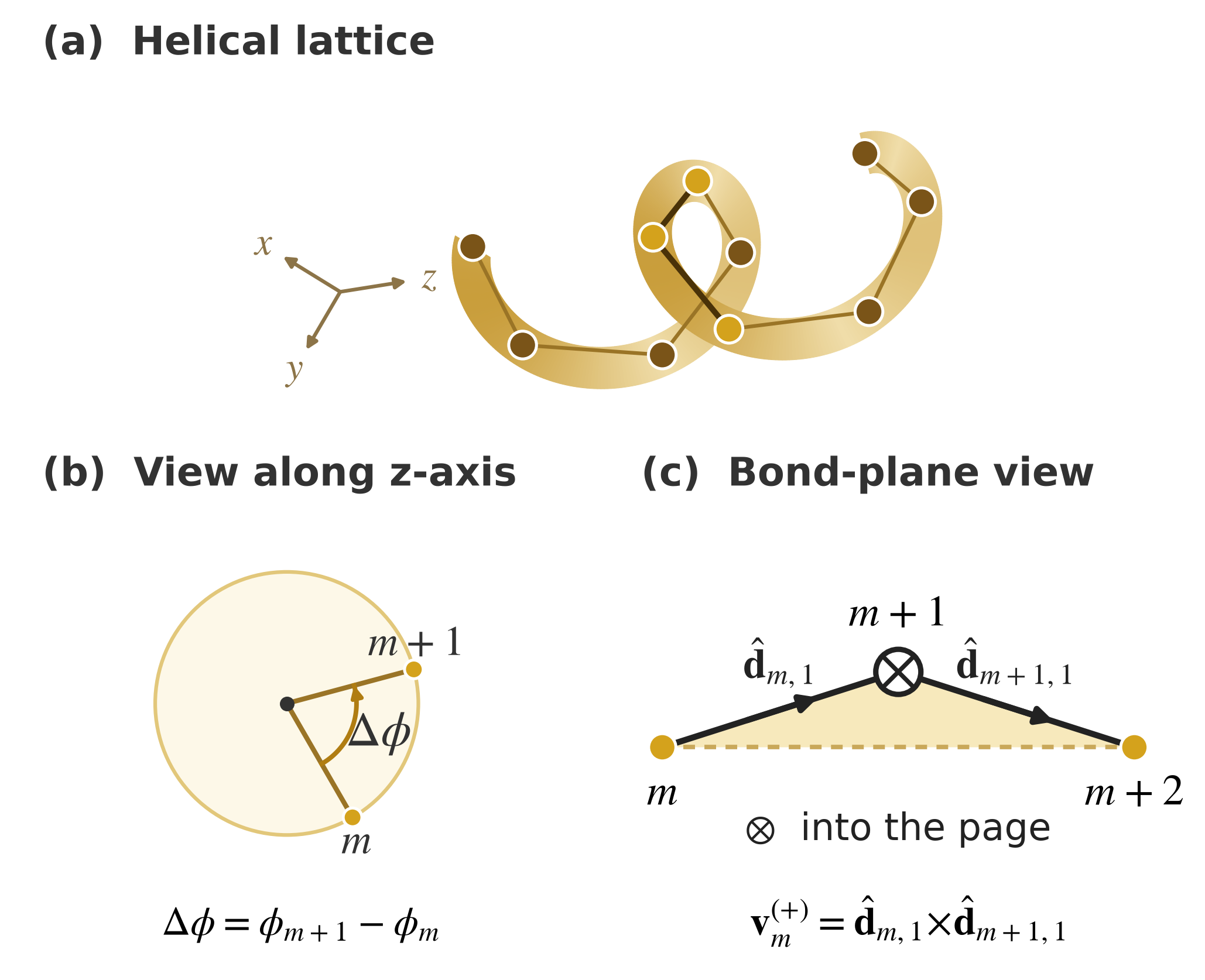}
    \caption{Geometry of the helical tight-binding model. (a) Ten-site helix with one spatial orbital per site. (b) Viewed along the helix axis, adjacent sites differ in azimuth by \(\Delta\phi=\phi_{m+1}-\phi_m\). (c) The cross product of successive unit bond vectors, \(\mathbf v_m^{(+)}=\hat{\mathbf d}_{m,1}\times\hat{\mathbf d}_{m+1,1}\), points into the page (\(\otimes\)) and enters the spin–orbit hopping term between sites \(m\) and \(m+2\) (dashed line).}
    \label{fig:helix-schematic}
\end{figure}

\subsection{Chiral tight-binding model}

Following Fransson, we use a tight-binding lattice model to represent the system.\cite{fransson_vibrational_2020} Consider a helix with $M$ sites at positions $\vec{r}_1,\dots,\vec{r}_n$ (see Figure \ref{fig:helix-schematic}). For a helix of radius $a$, axial extent ($z$-span) $c$, and $m_l$ loops, the positions of the sites (in Cartesian coordinates) are given by 
\begin{equation}
\vec{r}_m = \left(a\cos\phi_m, a\sin\phi_m, \frac{(m-1)c}{M-1}\right)
\end{equation}
where 
\begin{equation}
\phi_m = \frac{2\pi (m-1)m_l}{M-1}
\end{equation}
is the azimuthal angle of site $m$.
We define {\em bond vectors} connecting sites $m$ and $m+s$ by 
\begin{equation}
\hat{d}_{m,s} = \frac{\vec{r}_{m+s}-\vec{r}_m}{||\vec{r}_{m+s}-\vec{r}_m||}
\end{equation}
and the {\em chirality vector} by
\begin{equation}
\vec{v}_m^{(+)} = \hat{d}_{m,1}\times \hat{d}_{m+1,1}.
\end{equation}

The system Hamiltonian contains three terms corresponding to on-site energy, nearest neighbor hopping, and next-nearest neighbor SOC. Correspondingly, we define three system operators as \cite{fransson_vibrational_2020}
\begin{equation}
\begin{aligned}
    \hat{S}_1 &= \sum_{m=1}^{M} \hat{\psi}_m^\dagger \hat{\psi}_m\\
    \hat{S}_2 &= -\sum_{m=1}^{M-1} \hat{\psi}_m^\dagger \hat{\psi}_{m+1} + \text{h.c.}\\
    \hat S_3 & = \sum_{m=1}^{M-2} i\,\hat\psi_m^\dagger\, \vec v^{(+)}_m\cdot\hat{\boldsymbol\sigma}\,\hat\psi_{m+2} + \mathrm{h.c.}
\end{aligned}
\end{equation}
where $\hat{\psi}_m$ is the annihilation operator for an electron at site $m$ and $\hat{\sigma}$ is a vector of Pauli matrices. Concretely, $\hat{\psi}_m$ is a two-component vector operator 
\begin{equation}
\hat{\psi}_m = \begin{pmatrix}
    \hat{c}_{\uparrow,m}\\
    \hat{c}_{\downarrow,m}
\end{pmatrix}
\end{equation}
where $\hat{c}_{\uparrow,m}$ is the annihilation operator for a spin-up electron at site $m$ and $\hat{c}_{\downarrow,m}$ is the annihilation operator for a spin-down electron at site $m$. These operators obey the fermionic anticommutation relations
$\{\hat{c}_{s,m},\hat{c}_{s',m'}^\dagger\}
=\delta_{ss'}\delta_{mm'}$ and
$\{\hat{c}_{s,m},\hat{c}_{s',m'}\}=0$.
In the model used here, the SOC connects next-nearest neighbors
because $\vec{v}_m^{(+)}$ is constructed from the two successive
bonds along the path $m\to m+1\to m+2$, with the spin-dependent
hopping connecting the endpoints of that path.\cite{fransson_vibrational_2020}
The system Hamiltonian is then given by 
\begin{equation}\label{system}
\hat{\mathcal{H}}_S = \varepsilon_0 \hat{S}_1 + t_0 \hat{S}_2 + \lambda_0 \hat{S}_3
\end{equation}
where the terms $\varepsilon_0$, $t_0$, and $\lambda_0$ capture the magnitude of on-site energy, nearest-neighbor hopping, and next-nearest neighbor SOC, respectively. We restrict the calculation to the single-electron sector, for which the Hilbert space has dimension $2M$. In this sector, $\hat{S}_1$ is the identity operator, so that $\varepsilon_0$ represents a global shift in the system Hamiltonian $\hat{\mathcal{H}}_S$ and is set to zero. 

Due to the time-reversal (Kramers) symmetry of $\hat{\mathcal{H}}_S$, each energy level of the system Hamiltonian is doubly degenerate. In the site and spin-$z$ basis, the time-reversal operator is
$\Theta=(I_M\otimes i\sigma_y)K$, where $K$ denotes complex
conjugation. Time reversal changes the signs of both $i$ and
$\hat{\boldsymbol{\sigma}}$. Since the bond vectors are real,
$i\vec v_m^{(+)}\cdot\hat{\boldsymbol{\sigma}}$ is invariant,
as are the spin-independent on-site and hopping terms. Thus
$\Theta\hat{\mathcal H}_S\Theta^{-1}=\hat{\mathcal H}_S$ and since $\Theta^2=-I$, the single-electron eigenstates occur in
Kramers pairs.

To practically address this symmetry in our calculations without affecting the underlying physics, we add a small symmetry-breaking perturbation of the form $\delta \hat{\sigma}_{z}$ with $\delta = 10^{-8}\text{ cm}^{-1}$. It was verified that the computed decoherence rates and geometry-dependent trends are insensitive (i.e. vary by less than $0.01\%$) to the value of $\delta$ across several orders of magnitude.

We assume the bath (corresponding to vibrations of the nuclear degrees of freedom of the helix as well as host or solvent-induced degrees of freedom) consists of phonon modes of frequencies $\{ \omega_v \}$. Letting $\hat{a}_v$ denote the annihilation operator for the mode of frequency $\omega_v$, the bath Hamiltonian is given by 
\begin{equation}
\hat{\mathcal{H}}_B = \sum_{v=1}^{N_B} \hbar\omega_v \hat{a}_v^\dagger \hat{a}_v.
\end{equation}
Here $N_B$ is the number of bath modes. We make the assumption that $N_B$ is extremely large to justify the use of a smooth spectral density function to represent system-bath coupling.

\subsection{System--bath coupling}
In this minimal model, we conceptualize system-bath couplings as arising from bath-induced modulations of the parameters $\varepsilon_0$, $t_0$, and $\lambda_0$. To the first order, these modulations depend linearly on bath position operators ${ \hat{x}_v = \hat{a}_v + \hat{a}_v^\dagger }$ (corresponding to normal mode displacements whose fluctuations perturb electronic parameters). In Fransson's original model, system-bath coupling was constructed as\cite{fransson_vibrational_2020} 
\begin{equation}
\hat{\mathcal{H}}_{SB} = \hat{S}_1 \otimes \hat{B}_1 + \hat{S}_2\otimes \hat{B}_2 + \hat{S}_3 \otimes \hat{B}_3.
\end{equation}
where
\begin{equation}
\begin{aligned}
    \hat{B}_1 & = \sum_{v} \varepsilon_v (\hat{a}_v + \hat{a}_v^\dagger) \\
    \hat{B}_2 & = \sum_{v} t_v (\hat{a}_v + \hat{a}_v^\dagger) \\
    \hat{B}_3 & = \sum_{v} \lambda_v (\hat{a}_v + \hat{a}_v^\dagger).
\end{aligned}
\end{equation}
In particular, this model assumes a fully collective limit in which a single set of bath operators modulates amplitudes uniformly along the helix.\cite{upadhyay_weak_2026}

While this approximation is useful when considering electron transport, it is important to consider the role of local modulations along the helix when considering decoherence. We therefore utilize the general local coupling 
\begin{equation}
\begin{aligned}
    \hat{\mathcal{H}}_{SB}=\sum_{m}\hat{n}_m\otimes \hat{B}_m^\varepsilon + \sum_{m} \hat{T}_m \otimes \hat{B}^t_m + \sum_{m} \hat{L}_m\otimes \hat{B}_m^\lambda
\end{aligned}
\end{equation}
with per-site and per-bond operators 
\begin{equation}\label{site-operators}
\begin{aligned}
    \hat{n}_m &= \hat{\psi}_m^\dagger\hat{\psi}_m\\
    \hat{T}_m &= -(\hat{\psi}_m^\dagger\hat{\psi}_{m+1}+\text{h.c.})\\
    \hat L_m & = i\,\hat\psi_m^\dagger\, \vec v^{(+)}_m\cdot\hat{\boldsymbol\sigma}\,\hat\psi_{m+2} + \mathrm{h.c.}
\end{aligned}
\end{equation}
For compactness, we write $\hat S_m^\varepsilon=\hat n_m$,
$\hat S_m^t=\hat T_m$, and $\hat S_m^\lambda=\hat L_m$.
The allowed values of $m$ are $1,\ldots,M$ for the on-site
channel, $1,\ldots,M-1$ for hopping, and $1,\ldots,M-2$ for SOC.
Each sum over $m$ uses the range appropriate to its channel. The bath operators 
\begin{equation}\label{site-bath}
\hat{B}_m^c = \sum_{\nu} g_{m\nu}^c \hat{x}_\nu
\end{equation}
defined for site $m$ and channel $c$ (where $c$ represents either on-site energy ($c=\epsilon$), 
nearest neighbor hopping ($c=t$), 
or next nearest neighbor coupling ($c=\lambda$)) are also site-resolved. 
We specify the correlation structure of these operators in the next section. The globally correlated model used by Fransson can then be recovered as one limit of this generalized model. 

\subsection{Bath correlation functions and spectral densities} 
Using the site-resolved picture, the equilibrium correlation function of bath operators $\hat{B}_m^c$ and $\hat{B}_{m'}^{c'}$
is given by:\cite{nitzan_chemical_2024} 
\begin{equation}
\begin{aligned}
\langle \hat{B}_m^c(t)\hat{B}_{m'}^{c'}(0)\rangle
&= \sum_{\nu}g_{m\nu}^c g_{m'\nu}^{c'} \\
&\quad\times
\left[
\coth\left(\frac{\beta\hbar\omega_\nu}{2}\right)
\cos\omega_\nu t
-i\sin\omega_\nu t
\right].
\end{aligned}
\end{equation}
We define the spectral density tensor $\mathbf{J}$ to have components
\begin{equation}
J_{mm'}^{cc'}(\omega) = \pi \sum_{\nu} g_{m\nu}^c g_{m'\nu}^{c'}\delta(\omega-\omega_\nu)
\end{equation}
so that 
\begin{equation}\label{correlation-func}
\begin{aligned}
\langle \hat{B}_m^c(t)\hat{B}_{m'}^{c'}(0)\rangle
&= \frac{1}{\pi}\int_0^\infty d\omega\,
J_{mm'}^{cc'}(\omega) \\
&\quad\times
\left[
\coth\left(\frac{\beta\hbar\omega}{2}\right)
\cos\omega t
-i\sin\omega t
\right].
\end{aligned}
\end{equation}
For the sake of concreteness, and in order to account for 
correlations between bath-induced effects, we assume that $J_{mm'}^{cc'}(\omega)$ is Ohmic with exponential cutoff, with the prefactors that govern channel and spatial correlations defined as follows:
\begin{equation}\label{spectral-density}
J_{mm'}^{cc'}(\omega) =\underbrace{K_{cc'}\sqrt{A_cA_{c'}}}_{\text{channel}} \ \underbrace{F_{mm'}}_\text{spatial} \ \underbrace{\omega e^{-\omega/\omega_c}}_\text{Ohmic kernel}.
\end{equation}
Here $A_c$ dictates the strength of channel $c$ ($c= \varepsilon, t,\lambda$).  
The term 
\begin{equation}
K_{cc'} = \begin{cases}
    1 & c = c'\\
    \alpha & c \neq c'
\end{cases}
\end{equation}
dictates the correlation between different channels (where we assume that $\alpha$ is independent of $cc'$ for the sake of simplicity), and $F_{mm'}$ is the spatial correlation between different helix sites (which is assumed to decay exponentially, also for the sake of simplicity):
\begin{equation}
F_{mm'} = \exp(-|m-m'|\ell/\xi)~~.
\end{equation}
Here, $\xi/\ell$ is the correlation length (normalized by the fixed bond length, $l$). The limit $\xi/\ell\to \infty$ corresponds to a bath that is fully correlated across spatial sites while the limit $\xi/\ell\to 0$ corresponds to independent bonds. Note that Fransson's original model is recovered in the limit $\alpha = 1$, $\xi/\ell \to \infty$.

We assume $A_\varepsilon = 0$ 
since we are primarily interested in uncovering qualitative, geometry-dependent trends. We have verified that the sign of the geometric response in each limiting bath is unchanged for $0 \le A_\varepsilon/A_t \le 100$, and that the collective baths are unaffected by $A_\varepsilon$.

Furthermore, the relative amplitudes $A_t$ and $A_\lambda$ are not fully independent in a given physical system. Because $A_t$ and $A_\lambda$ characterize the spectral amplitudes of fluctuations of $t$ and $\lambda$, respectively, we can relate them through corresponding root-mean square (RMS) fractional fluctuations. For channel $c\in \{t,\lambda\}$ and site $m$, we have from Eq. (\ref{correlation-func}):
\begin{equation}
\begin{aligned}
    \langle \hat{B}_m^c(0)^2\rangle &= \frac{1}{\pi}\int_0^\infty d\omega J_{mm}^{cc}(\omega)\coth\left(\frac{\beta \hbar \omega}{2}\right) \\& = \frac{A_c}{\pi}\int_0^\infty d\omega \omega e^{-\omega/\omega_c}\coth\left(\frac{\beta \hbar \omega}{2}\right)
\end{aligned}
\end{equation}
so that 
\begin{equation}
    \frac{\sqrt{\langle \hat{B}_m^\lambda(0)^2\rangle}}{\sqrt{\langle \hat{B}_m^t(0)^2\rangle}}  = \sqrt{\frac{A_\lambda}{A_t}}.
\end{equation}
Now, let the relative fractional fluctuation amplitude be defined by 
\begin{equation}\label{frac-sens}
    \begin{aligned}
        \eta &= \frac{\sqrt{\langle \hat{B}_m^\lambda(0)^2\rangle}/\lambda_0}{\sqrt{\langle \hat{B}_m^t(0)^2\rangle}/t_0} \\& = \frac{t_0}{\lambda_0}\sqrt{\frac{A_\lambda}{A_t}}~~.
    \end{aligned}
\end{equation}
In the absence of microscopic information indicating that either interaction is fractionally more sensitive to environmental perturbations, we assume that $\eta = 1$ as a reference value, which is equivalen to assuming that the fractional fluctuation amplitudes in the two channels are comparable. Under this simplifying assumption, Eq. (\ref{frac-sens}) reduces to: 
\begin{equation}\label{frac-sens-2}
    \frac{\lambda_0}{t_0} = \sqrt{\frac{A_\lambda}{A_t}}.
\end{equation}
We have verified that the achiral maximum in the local, uncorrelated bath and the achiral minimum in the collective, correlated bath persist for $0.5 \le \eta \le 2$, although the extent of the inversion region in Fig.~4 grows with $\eta$.

\subsection{Redfield dynamics and decoherence rates}
Projecting the system-bath coupling onto the system energy eigenbasis $\{\ket{i}\}$, such that $\hat{\mathcal{H}}_S \ket{i} = E_i \ket{i}$,
yields 
\begin{equation}
\begin{aligned}
    \hat{\mathcal{H}}_{SB} & = \sum_{i,j} \ket{i}\bra{i}\left[\sum_{m}\left(\hat{n}_m\otimes \hat{B}_m^\varepsilon + \hat{T}_m \otimes \hat{B}^t_m + \hat{L}_m\otimes \hat{B}_m^\lambda \right)\right]\ket{j}\bra{j} \\& = \sum_{i,j}\ket{i}\bra{j} \otimes \hat{\Lambda}_{ij}.
\end{aligned}
\end{equation}
where 
\begin{equation}
\hat{\Lambda}_{ij} = \sum_{m}\left(\braket{i|\hat{n}_m|j}\hat{B}_m^\varepsilon+\braket{i|\hat{T}_m|j}\hat{B}_m^t+\braket{i|\hat{L}_m|j}\hat{B}_m^\lambda\right).
\end{equation}
The $\hat{\Lambda}_{ij}$ correlation functions are then given by 
\begin{equation}\label{cor-func}
\begin{aligned}
    C_{ij,mn}(t) &= \langle \hat{\Lambda}_{ij}(t)\hat{\Lambda}_{mn} \rangle \\& = \sum_{p,q}\sum_{c,c'\in\{\varepsilon,t,\lambda\}}
    \braket{i|\hat{S}_p^c|j}
    \braket{m|\hat{S}_q^{c'}|n}
    \left\langle \hat{B}_p^c(t)\hat{B}_q^{c'}\right\rangle
\end{aligned}
\end{equation}
where the sums are over pairs of lattice sites $(p,q)$ and pairs of decoherence channels $(c,c').$

Under the assumption of weak system-bath coupling (second order perturbation theory with the system-bath coupling as the perturbation) and Markovianity (the assumption that the bath correlation time is much shorter than the system relaxation time), the time evolution of the system reduced density operator $\hat{\sigma} (t)$, in the energy eigenbasis $\{\ket{i}\}$, is given by the non-secular Redfield quantum master equation:\cite{nitzan_chemical_2024}
\begin{equation}\label{Redfield}
\begin{aligned}
    \frac{d}{dt}\sigma_{jk}(t) & = -i\omega_{jk}\sigma_{jk}(t) - \sum_{m,n}\Gamma_{jn,nm}(\omega_{mn}) \sigma_{mk}(t) \\
    & \quad - \sum_{m,n}\Gamma^*_{km,mn}\sigma_{jn}(t)+\sum_{m,n}\Gamma_{nk,jm}(\omega_{mj})\sigma_{mn}(t)  \\& \quad + \sum_{m,n}\Gamma_{mj,kn}^*(\omega_{nk})\sigma_{mn}(t).
\end{aligned}
\end{equation}
Here $\omega_{mn} = (E_m - E_n)/\hbar$ is the Bohr frequency 
of the $m\to n$ transition between system energy states and 
\begin{equation}
\begin{aligned}
    \Gamma_{jk,mn}(\omega) &= \frac1{\hbar^2}\int_0^\infty d\tau e^{i\omega \tau} C_{jk,mn}(\tau) \\& \approx \frac1{\hbar^2}\text{Re}\int_0^\infty d\tau e^{i\omega \tau} C_{jk,mn}(\tau).
\end{aligned}
\end{equation}
We have neglected the imaginary part of the above integral because the imaginary part contributes mainly to oscillatory dynamics while the real part contributes primarily to decoherence dynamics, the main focus of this work (strictly speaking, this statement is true in the secular approximation which we describe below). 

In this work, we are primarily interested in the decoherence of a prepared superposition between Kramers doublets. We order the single-electron eigenstates by increasing energy,
starting with $i=0$. The states $\ket{0}$ and $\ket{2}$ select one
representative coherence associated with the two lowest Kramers doublets.
We use their superposition as a low-energy probe of interdoublet
coherence and prepare the initial pure state
\begin{equation}\label{superposition}
    \hat\sigma(0)=\ket{\Psi}\bra{\Psi}, \qquad
    \ket{\Psi}=\frac{\ket{0}+\ket{2}}{\sqrt{2}} .
\end{equation}
We examine the decay of $\sigma_{02}(t)=\braket{0|\hat\sigma(t)|2}$,
for which $\sigma_{02}(0)=1/2$. $\delta\hat\sigma_z$ lifts the exact degeneracy within each Kramers doublet and thereby fixes the choice of basis within each doublet.

Within Redfield theory, the weak system-bath coupling and Markovian approximations are often accompanied by the secular approximation. Within the secular approximation, populations only couple to other populations and coherences to evolve independently.\cite{nitzan_chemical_2024} The standard justification for the secular approximation starts from Eq. (\ref{Redfield}), which we rewrite in standard Redfield form:
\begin{equation}
    \frac{d}{dt}\sigma_{jk}(t)
    =
    -i\omega_{jk}\sigma_{jk}(t)
    +
    \sum_{m,n} R_{jk,mn}\sigma_{mn}(t)
\end{equation}
with 
\begin{equation}\label{redfield-coef}
    \begin{aligned}
    R_{jk,mn} & = -\delta_{kn}\sum_{l}\Gamma_{jl,lm}(\omega_{ml}) - \delta_{jm}\sum_{l}\Gamma^*_{kl,ln}(\omega_{nl}) \\& \quad + \Gamma_{nk,jm}(\omega_{mj})+\Gamma^*_{mj,kn}(\omega_{nk}).
\end{aligned}
\end{equation}
Moving to the interaction picture and defining
\begin{equation}
    \tilde{\sigma}_{jk}(t) = e^{i\omega_{jk}t}\sigma_{jk}(t)
\end{equation}
we obtain 
\begin{equation}
    \frac{d}{dt}\tilde{\sigma}_{jk}(t)
    =
    \sum_{m,n}
    R_{jk,mn}
    e^{i(\omega_{jk}-\omega_{mn})t}
    \tilde{\sigma}_{mn}(t).
\end{equation}
In particular, the phase factor $e^{i(\omega_{jk}-\omega_{mn})}$ is assumed to oscillate rapidly if 
$\omega_{jk}-\omega_{mn}$ is large (compared to the relevant system decoherence rates). Thus, terms containing $R_{jk,mn}e^{i(\omega_{jk}-\omega_{mn}) t}$ average to zero 
except for terms where $mn=jk$. 
Under the secular approximation, the decoherence rate of the $\sigma_{02}$ coherence is given by:
(see Eq. \ref{redfield-coef})  
\begin{equation}\label{eq:decoherence_split}
\begin{aligned}
\Gamma_{02} &= -R_{02,02}
\\&= \sum_{l\neq0}\Gamma_{0l,l0}(\omega_{0l}) + \sum_{l\neq2}\Gamma^*_{2l,l2}(\omega_{2l})
\\&\quad + \Gamma_{00,00}(0) + \Gamma^*_{22,22}(0) - \Gamma_{22,00}(0) - \Gamma^*_{00,22}(0)
\end{aligned}
\end{equation}
Furthermore, the overall decoherence rate can be decomposed as follows:  
\begin{equation}\label{decoherence-split}
\Gamma_{02} = \Gamma_{02}^{\mathrm{rel}} +\Gamma_{02}^{\phi}.
\end{equation}
where  
\begin{equation}
\begin{aligned}
\Gamma^{\mathrm{rel}}_{02} &= \sum_{l\neq0}\Gamma_{0l,l0}(\omega_{0l}) + \sum_{l\neq2}\Gamma^*_{2l,l2}(\omega_{2l})
\end{aligned}
\end{equation}
is the population-relaxation induced contribution to decoherence and  \begin{equation}
\begin{aligned}\Gamma^{\phi}_{02} &= \Gamma_{00,00}(0) + \Gamma^*_{22,22}(0) \\
& \quad - \Gamma_{22,00}(0) - \Gamma^*_{00,22}(0)\end{aligned}\end{equation}
is the pure dephasing contribution to decoherence.\cite{nitzan_chemical_2024}

Because the Kramers pairs give rise to nearly coincident transition
frequencies, we compared the diagonal secular rate with the nonsecular
Redfield decay rate of $\sigma_{02}$. The comparison included all four
limiting combinations of spatial and channel correlations, sampled twist
angles from $30^\circ$ to $330^\circ$, and SOC strengths
$\lambda_0/t_0=0.05,\,0.15,\,0.30,\,0.50,\,0.75$.
Across the points tested, the relative discrepancy
$\lvert\Gamma_{02}^{\rm nonsec}/\Gamma_{02}^{\rm sec}-1\rvert$
was below $0.02\%$ for the real bath kernel and below $0.2\%$ for the
full complex kernel, both relative to the same secular rate. 
We therefore use the secular rate $\Gamma_{02}^{\rm sec}=-\operatorname{Re}R_{02,02}$
for the parameter scans.

\subsection{Parameter regimes}
As a minimal and computationally tractable model, we present results for a helix with $M =10$ sites. We later show that the sign and location of the geometric extrema are robust to helix length. For the sake of concreteness, we also choose to focus on a family of helices with a fixed ratio of rise per bond to bond length, $h/\ell = \sqrt{3}/2$, so that the achiral geometry ($\Delta\phi = 180^\circ$) is a planar zig-zag with $120^\circ$ bond angles.

We choose $t_0 = 0.15\text{ eV}$, which is representative of electronic transfer energies in strongly coupled molecular systems, and use this value as a reference energy scale.\cite{fransson_vibrational_2020, chiesa_many-body_2024} We then characterize SOC strength via the dimensionless ratio $\lambda_0/t_0$. In particular, we explore the behavior of the model in three regimes, a weak-SOC regime corresponding to $\lambda_0/t_0 \lesssim 0.05$, an intermediate-SOC regime corresponding to $0.05\lesssim\lambda_0/t_0\lesssim 0.3$, and a strong-SOC regime corresponding to $\lambda_0/t_0\gtrsim 0.3$. These designations serve only to distinguish regimes of the model and are not intended as universal or material-specific boundaries. More broadly, the minimal model employed here is designed to isolate the interplay between helical geometry and SOC in electronic decoherence. Accordingly, our results should be interpreted as predictions of qualitative physical trends rather than quantitative predictions for a specific material.

Furthermore, we fix $A_t/\hbar^2 = 9.0\times 10^{-4}$ ($A_t$ has the same dimensions as $\hbar^2$ so the resulting ratio is dimensionless) and $A_\lambda$ is given by Eq. (\ref{frac-sens-2}). This value was chosen to ensure that the Born-Markov conditions for Redfield theory are met. Furthermore, we employ a spectral cutoff of $\omega_c = 500 \text{ cm}^{-1}$ for the Ohmic spectral density (\ref{spectral-density}) and all calculations are carried out at $T=300 \text{ K}.$

\subsection{Geometry scan}

Our primary objective is to evaluate the extent to which variations in a system's chiral morphology result in modulations of electronic decoherence rates and how this relationship between geometry and decoherence is related to SOC strength and bath correlations. Since we are taking a helical lattice as our paradigmatic model, we represent the degree of twist by the azimuthal step between adjacent sites, denoted as $\Delta \phi$. We then consider decoherence rates across a family of helices of fixed length as we continuously vary $\Delta \phi$ (thereby continuously varying the helical morphology). Such a scan may be interpreted either as a single system with a continuously tunable twist or as a family of structurally related systems whose principal difference is their helical geometry. To isolate the effect of geometry from changes in the nearest-neighbor hopping amplitude, we hold the nearest-neighbor bond length \(\ell\) fixed throughout the scan. Consequently, the helix radius is allowed to vary with \(\Delta\phi\). Because the radius grows as $a = \ell/[4\sin(\Delta\phi/2)]$ at fixed bond length, it exceeds the bond length for $\Delta\phi \lesssim 29^\circ$, where the structure is a shallow arc rather than a helix. We therefore restrict all scans to $30^\circ \le \Delta\phi \le 330^\circ$.

\section{Results and Discussion}

\subsection{Geometric control requires intermediate and strong SOC}
\begin{figure*}
    \centering
    \includegraphics[width=1.0\textwidth]{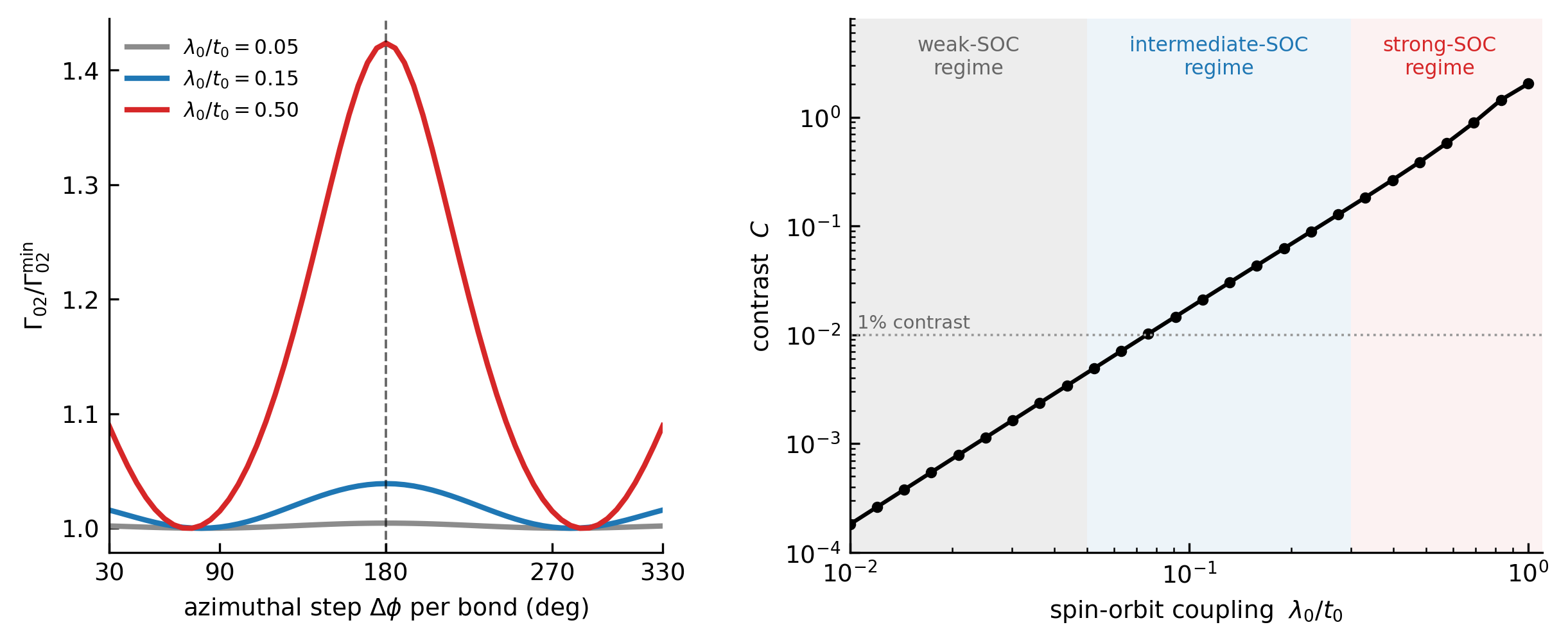}
    \caption{(a) A plot of normalized decoherence rates $\Gamma_{02}/\Gamma_{02}^\text{min}$ as a function of azimuthal step $\Delta \phi$
    for an uncorrelated bath ($\alpha = 0$, $\xi/\ell \to 0$). Geometric control is negligible in the weak SOC regime ($\lambda_0/t_0 = 0.05$; grey line), observable in the intermediate SOC regime ($\lambda_0/t_0=0.15$; blue line), and highly pronounced in the strong SOC regime ($\lambda_0/t_0 = 0.5$; red line). (b) Plot of geometric contrast $\Gamma_{02}^\text{max}/\Gamma_{02}^\text{min}-1$ as a function of SOC strength ($\alpha = 0$, $\xi/\ell \to 0$). The geometric contrast grows with SOC strength $\lambda_0/t_0$ and a contrast of $>1\%$ is observable only in the intermediate- and strong-SOC regimes.}
    \label{fig:geom_control_1}
\end{figure*}

Because system chirality enters the system Hamiltonian, Eq. (\ref{system}), through $\hat{S}_3$, we expect that the extent of geometric control will be modulated by the SOC strength $\lambda_0/t_0$. This is reflected in Figure \ref{fig:geom_control_1}(a) which, for an uncorrelated bath ($\alpha = 0$, $\xi/\ell \to 0$), plots the normalized decoherence rate $\Gamma_{02}/\Gamma_{02}^\text{min}$ as the azimuthal step (twist) $\Delta \phi$ varies from $30^\circ$ to $330^\circ$ for three SOC strengths $\lambda_0/t_0 = 0.05, 0.15, 0.50$ corresponding to the weak-, intermediate-, and strong-SOC regimes, respectively. $\Gamma_{02}^\text{min}$ is the minimum absolute decoherence rate across the scanned values of $\Delta \phi$ and is used to normalize $\Gamma_{02}$ so that we isolate the geometry-dependent trend in $\Gamma_{02}$. We note that the dependence of $\Gamma_{02}/\Gamma_{02}^\text{min}$ on $\Delta \phi$ is negligible for the weak-SOC case $\lambda_0/t_0 = 0.05$, observable--albeit weak-- in the intermediate-SOC case $\lambda_0/t_0 = 0.15$, and substantial in the strong-SOC case $\lambda_0/t_0 = 0.50.$

To further quantify the strength of the geometry-dependence, we define the geometric contrast by 
\begin{equation}
    C = \frac{\Gamma_{02}^\text{max}-\Gamma_{02}^\text{min}}{\Gamma_{02}^\text{min}}= \frac{\Gamma_{02}^\text{max}}{\Gamma_{02}^\text{min}}-1
\end{equation}
to reflect the percentage difference between the minimum and maximum decoherence rates. Figure \ref{fig:geom_control_1}(b) shows the variation of $C$ with SOC strength $\lambda_0/t_0$. As expected, the geometric contrast grows monotonically with SOC strength. Of particular note is the fact that a geometric contrast of $1\%$ or greater is only achieved in the intermediate-and high-SOC regimes. Therefore, the geometric control effect presented here is a characteristic of intermediate- and high-SOC systems, but not of low-SOC systems (e.g. unmodified organic helices). 

To fully appreciate the implications of the geometric control observed in Figure \ref{fig:geom_control_1}, it is instructive to examine the specific helical morphologies corresponding to the extrema in the plot. In particular, the normalized decoherence rate $\Gamma_{02}/\Gamma_{02}^{\mathrm{min}}$ reaches a maximum at $\Delta\phi = 180^\circ$, corresponding to a flat, planar zig-zag (the achiral limiting case of the lattice). Conversely, the decoherence rate is minimized at $\Delta\phi \approx 73$--$82^\circ$, depending on SOC strength, corresponding to coiled, chiral structures with roughly 4.4--4.9 sites per helical turn. Thus, for a helical lattice in an uncorrelated bath, chiral geometries decohere more slowly than the achiral planar state. As shown in Section \ref{pop}, this behavior has two reinforcing origins. An uncorrelated bath combines the local responses of the electronic energy gaps independently, so their contributions add without cancellation, and this total susceptibility is largest at the achiral geometry. Population relaxation, which contributes comparably to $\Gamma_{02}$, is also maximized there at the temperature considered.

\subsection{Bath correlations define the nature of geometric control}
\begin{figure}
    \centering
    \includegraphics[width=1.0\linewidth]{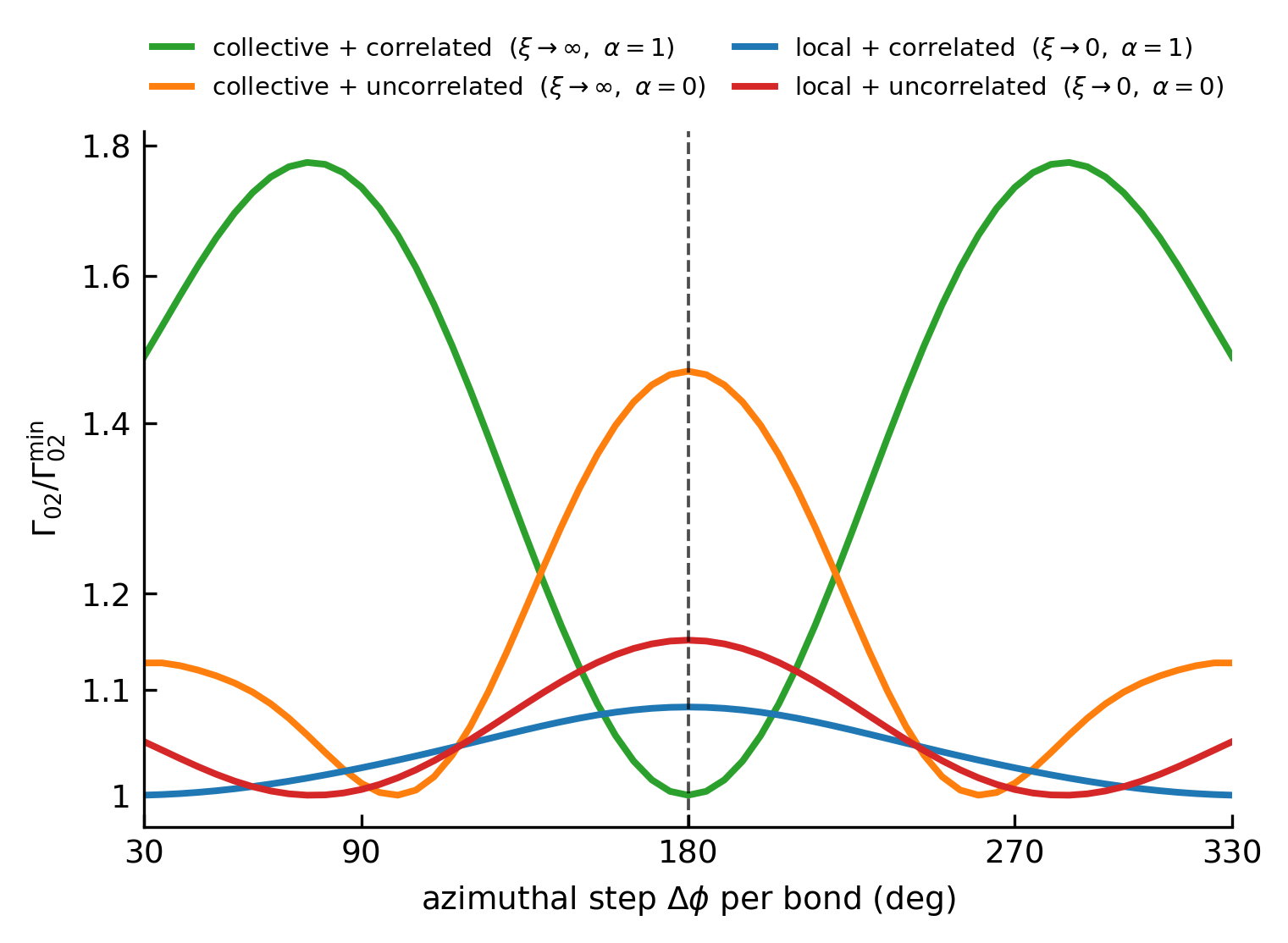}
    \caption{The normalized decoherence rate $\Gamma_{02}/\Gamma_{02}^\text{min}$ is plotted for four limiting bath correlation cases: $\alpha=0$, $\xi/\ell\to 0$ (red); $\alpha = 1$,  $\xi/\ell\to 0$ (blue); $\alpha = 0$, $\xi/\ell \to \infty$ (orange); and $\alpha = 1$, $\xi/\ell\to \infty$ (green).
    }
    \label{fig:correlations}
\end{figure}

The results in the previous section are presented in the case of a fully uncorrelated bath ($\alpha = 0$, $\xi/\ell \to 0$). Because our model consists of two separate correlation parameters, $\alpha$ (controlling the correlation between the hopping and SOC channels) and $\xi/\ell$ (controlling the correlation between sites on the helical lattice), we consider four limiting combinations of these parameters shown in Figure \ref{fig:correlations}. In particular, we notice that changing the correlation structure of the bath can modulate the overall geometry dependence of decoherence rates in our helical system. In the cases of a local and uncorrelated bath ($\alpha = 0$, $\xi/\ell \to 0$), local and channel-correlated bath ($\alpha = 1$, $\xi/\ell \to 0$), and spatially collective and channel uncorrelated bath ($\alpha = 0$,$\xi/\ell \to \infty$), we see that the normalized decoherence rate reaches a maximum at the achiral geometry. However, in the case of a collective, channel correlated bath $(\alpha = 1$, $\xi/\ell\to \infty$), we observe the opposite behavior where the achiral planar state minimizes the normalized decoherence rate relative to chiral geometries. Therefore, under some bath correlation structures, decoherence is suppressed at chiral geometries relative to achiral geometries whereas in others, decoherence is suppressed in achiral geometries. 

\begin{figure*}
    \centering
    \includegraphics[width=1\linewidth]{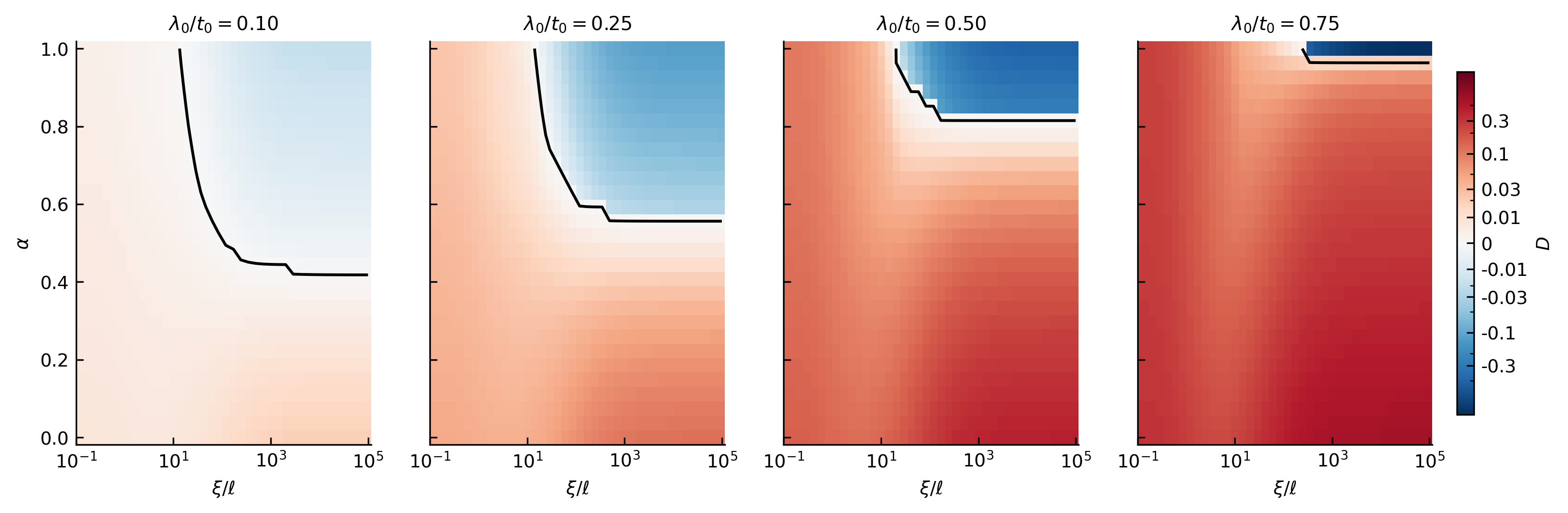}
    \caption{Map of the achiral-to-chiral decoherence ratio $D$ defined in Eq. (\ref{D}) across bath correlation parameter space for the SOC strengths $\lambda_0/t_0$ indicated above each panel ($\eta = 1$, $A_\varepsilon = 0$). Positive values indicate that the achiral zig-zag geometry decoheres more rapidly than the best chiral geometry, whereas negative values indicate that the achiral geometry decoheres more slowly. The black contour denotes the boundary separating these two regions.}
    \label{fig:phase diagram}
\end{figure*}

The limiting cases in Figure \ref{fig:correlations} demonstrate that bath correlations can invert geometric control, but real, physically realizable systems need not occupy either extreme. Rather, the bath correlation parameters $\alpha$ and $\xi/\ell$ can be considered to be members of a continuous configuration space. We map the geometry-decoherence relationship over this $(\alpha, \xi/\ell)$ space in Figure \ref{fig:phase diagram}. 

To quantify the degree to which decoherence is suppressed/enhanced at chiral geometries relative to achiral geometries, we define the quantity 
\begin{equation}\label{D}
    D = \log_{10}\frac{\Gamma_{02}(180^\circ)}{\Gamma_{02}(\Delta\phi^\star)}
\end{equation}
where $\Gamma_{02}(\Delta\phi^\star)$ is the lowest local minimum of $\Gamma_{02}$ over the chiral geometries $30^\circ \le \Delta\phi < 180^\circ$, with the endpoint $\Delta\phi = 30^\circ$ counted as a minimum when $\Gamma_{02}$ increases away from it. If $D > 0$, the achiral geometry decoheres more rapidly than the best chiral geometry. If $D < 0$, $\Gamma_{02}(180^\circ) < \Gamma_{02}(\Delta\phi)$ for all $\Delta\phi \in [30^\circ, 180^\circ)$, so the achiral geometry decoheres more slowly than every chiral geometry. The black $D = 0$ contour in Figure~\ref{fig:phase diagram} is the inversion boundary at which $\Gamma_{02}(180^\circ) = \Gamma_{02}(\Delta\phi^\star)$.

The results in Figure \ref{fig:phase diagram} reveals the physical principles governing our room-temperature system. At low $\alpha$ and/or short $\xi/\ell$, decoherence is reduced at chiral geometries relative to the achiral zig-zag geometry. The reduction in decoherence rate for achiral geometries relative to chiral geometries observed in Figure \ref{fig:correlations} only emerges in the case of simultaneously strong channel correlation and long spatial correlation length. However, it is important to note that this inversion does not arise only at the singular fully collective limit; for $\lambda_0/t_0 \le 0.5$, it occupies a finite region of environmental correlation space and thus may be relevant to certain highly correlated systems. The extent of this region depends on the relative fluctuation amplitudes: it grows with $\eta$ and nearly vanishes below $\eta = 0.5$, and it shrinks as local site-energy fluctuations increase, nearly vanishing for $A_\varepsilon/A_t \gtrsim 30$ (Figs.~S3 and S4). Achiral decoherence protection therefore additionally requires SOC fluctuations comparable to or larger than hopping fluctuations, in fractional terms, and weak local site-energy noise.

Taken together with Figure \ref{fig:geom_control_1}, these results reveal distinct roles for intrinsic and environmental parameters. In particular, increasing system SOC amplifies the sensitivity of the decoherence rate to changes in geometry and reduces the region of space in which achiral decoherence protection is observed. In addition, the relative fluctuation amplitudes $\eta$ and $A_\varepsilon$ further modulate the size of this region. On the other hand, bath correlations (which are a fixed property of the system-bath coupling) determine whether chirality suppresses or enhances decoherence relative to the achiral state for that particular system.
\subsection{The origin of bath correlation influence on geometric control} \label{pop}

\begin{figure*}
    \centering
    \includegraphics[width=1\linewidth]{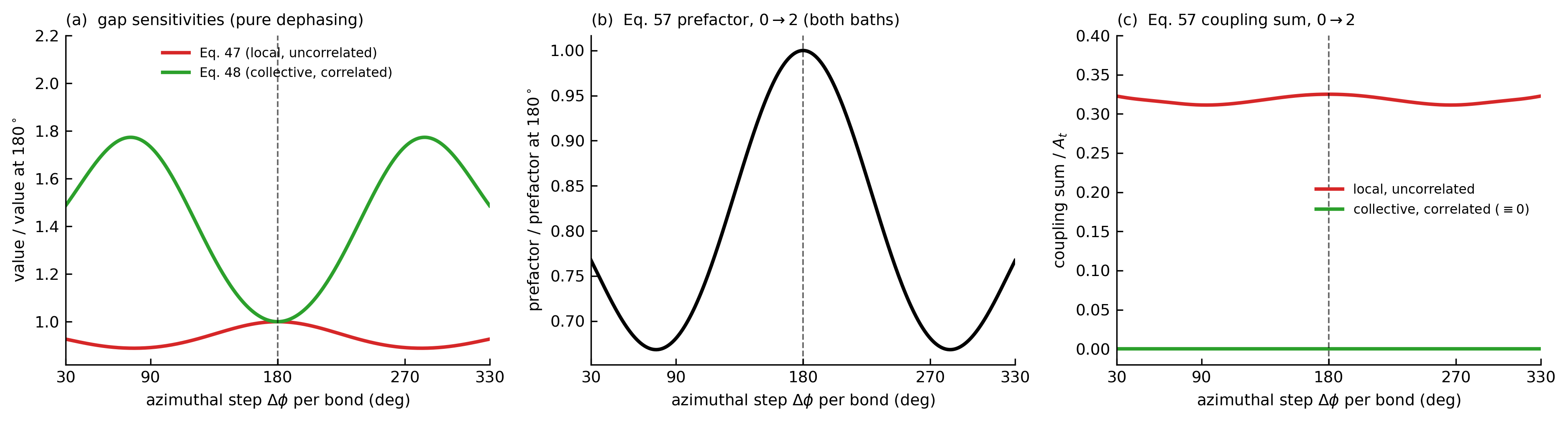}
    \caption{Geometric origin of the two decoherence contributions ($\lambda_0/t_0 = 0.3$).
(a) Pure-dephasing gap sensitivities, Eq.~(47) for the local, uncorrelated bath (red) and
Eq.~(48) for the collective, correlated bath (green), normalized to their values at
$\Delta\phi = 180^\circ$.
(b) Frequency prefactor of the $0\to2$ relaxation rate in Eq.~(57), common to both baths,
normalized to its value at $180^\circ$.
(c) Coupling sum of Eq.~(57) for $0\to2$ in units of $A_t$; it vanishes identically in the
collective, correlated bath.
Dashed lines mark the achiral geometry.}
    \label{fig:mechanism}
\end{figure*}

In order to understand the mechanism by which bath correlations influence the nature of geometric control over decoherence rates, we consider the two extreme cases of a collective, correlated bath ($\alpha = 1$, $\xi/\ell\to \infty$) and a local, uncorrelated bath ($\alpha = 0$, $\xi/\ell \to 0$). We return to Eq. (\ref{decoherence-split}) which splits the total decoherence rate into contributions from pure dephasing and population relaxation and consider the molecular basis of each contribution in turn.

Consider a state initially given by the superposition in Eq. (\ref{superposition}). Pure dephasing is the result of elastic interactions with the bath that do not induce transitions between states but perturb the relative phase between $\ket{0}$ and $\ket{2}$ in a random manner, thereby reducing the magnitude of the ensemble-averaged coherence.

Quantitatively, the bath-induced fluctuation of the $0$-$2$ energy gap is given by
\begin{equation}
    \delta \hat{\Delta}_{02}(t) = \hat{\Lambda}_{00}(t)-\hat{\Lambda}_{22}(t)~~,
\end{equation}
and the gap correlation function is then 
\begin{align}
    C_{\Delta,02}(t) = \langle \delta \hat{\Delta}_{02}(t)\delta \hat{\Delta}_{02}(0)\rangle~~,
\end{align}
so that the pure dephasing rate constant is \cite{kuhn_dynamics_2023}
\begin{align}
    \Gamma^\phi_{02} = \frac{1}{\hbar^2}\text{Re}\int_0^\infty dt C_{\Delta,02}(t).
\end{align}
For our model, we can define 
\begin{equation}
    \Delta S_{m}^c = \braket{0|\hat{S}_m^c|0} - \braket{2|\hat{S}_m^c|2}
\end{equation}
to be the sensitivity of the $0$-$2$ energy gap to a local fluctuation of channel $c$ at  site $m$ ($\hat{S}_m^c$ is defined as in (\ref{site-operators})). Because 
$\hat{\mathcal{H}}_S$
is explicitly geometry-dependent through the chirality vector $\vec{v}_m^{(+)}$, changes in geometry change the eigenstates and therefore the 
pattern of bath sensitivities. 
Specifically: 
\begin{align}
   C_{\Delta,02}(t) = \sum_{m,m'}\sum_{c,c'}\Delta S_m^c \Delta S_{m'}^{c'}\langle \hat{B}_m^c(t)\hat{B}_{m'}^{c'}(0)\rangle 
\end{align}
and 
\begin{equation}
    \begin{aligned}
        \Gamma_{02}^{\phi}
&=
\frac{1}{\hbar^2}
\operatorname{Re}
\sum_{m,m'}
\sum_{c,c'}
\Delta S_m^c
\Delta S_{m'}^{c'}
\int_0^\infty dt\,
\left\langle
\hat{B}_m^c(t)
\hat{B}_{m'}^{c'}(0)
\right\rangle 
\\[4pt]
&=
\frac{1}{2\hbar^2}
\sum_{m,m'}
\sum_{c,c'}
\Delta S_m^c
\Delta S_{m'}^{c'}
\lim_{\omega\to 0^+}
\left[
J_{mm'}^{cc'}(\omega)
\coth\left(
\frac{\beta\hbar\omega}{2}
\right)
\right]
    \end{aligned}
\end{equation}
where $\hat{B}_m^c$ is defined as in (\ref{site-bath}).
For our spectral density model under consideration here (see Eq. \eqref{spectral-density}), this becomes 
\begin{align}\label{pure-dephasing}
\Gamma_{02}^{\phi}
&=
\frac{1}{\beta\hbar^3}
\sum_{m,m'}
\sum_{c,c'}
\Delta S_m^c
\Delta S_{m'}^{c'}
K_{cc'}
\sqrt{A_c A_{c'}}
F_{mm'}
\end{align}
Eq. (\ref{pure-dephasing}) makes more transparent the role of bath correlations in influencing the decoherence rates. In particular it separates the electronic response, contained in the geometry-dependent quantities $\Delta S_{m}^c$ from the environmental correlation structure contained in the product $K_{cc'}F_{mm'}$.

In the fully uncorrelated limit { ($\alpha = 0$, $\xi/\ell \to 0$)}, fluctuations at a given site and channel carry no information about fluctuations at any other site or channel. As such, each local contribution to the gap fluctuation enters independently and is added quadratically (weighted by the corresponding channel amplitude):
\begin{equation}
    \Gamma_{02}^\phi \sim \sum_{m,c}A_c(\Delta S_{m}^c)^2
\end{equation}
Gap sensitivities thus cannot cancel each other out and can only add to the overall magnitude of the system's distributed susceptibility to local environmental perturbations. This total amplitude-weighted susceptibility is maximized at the achiral geometry, contributing to the achiral maximum in Figure \ref{fig:mechanism}a.

On the other hand, in the case that the bath is fully correlated {($\alpha = 1$, $\xi/\ell\to \infty$)}, the pure dephasing rate includes cross-terms and is given by the square of the amplitude weighted sum of gap sensitivities:
\begin{equation}
    \begin{aligned}
        \Gamma_{02}^\phi &\sim \sum_{m,m'}\sum_{c,c'}\sqrt{A_cA_{c'}}\Delta S_{m}^c\Delta S_{m'}^{c'} 
        = \left(\sum_{m,c}\sqrt{A_c}\Delta S_m^c\right)^2.
    \end{aligned}
\end{equation}
Because the collective perturbation acts simultaneously across sites and channels, the sensitivities may be summed in determining the overall system response. Because these sensitivities are signed quantities, local responses carrying opposite signs may cancel while responses carrying the same sign may reinforce each other. In this case, maximum cancellation occurs at the achiral geometry (see Figure \ref{fig:mechanism}a) so that decoherence is suppresed for that geometry relative to nearby chiral geometries. Bath correlations therefore change how geometry-dependent local responses are combined in determining the total pure dephasing rate. As a result, changing bath correlations can change which geometries are most or least susceptible to decoherence.  
\begin{figure*}
    \centering
    \includegraphics[width=1\linewidth]{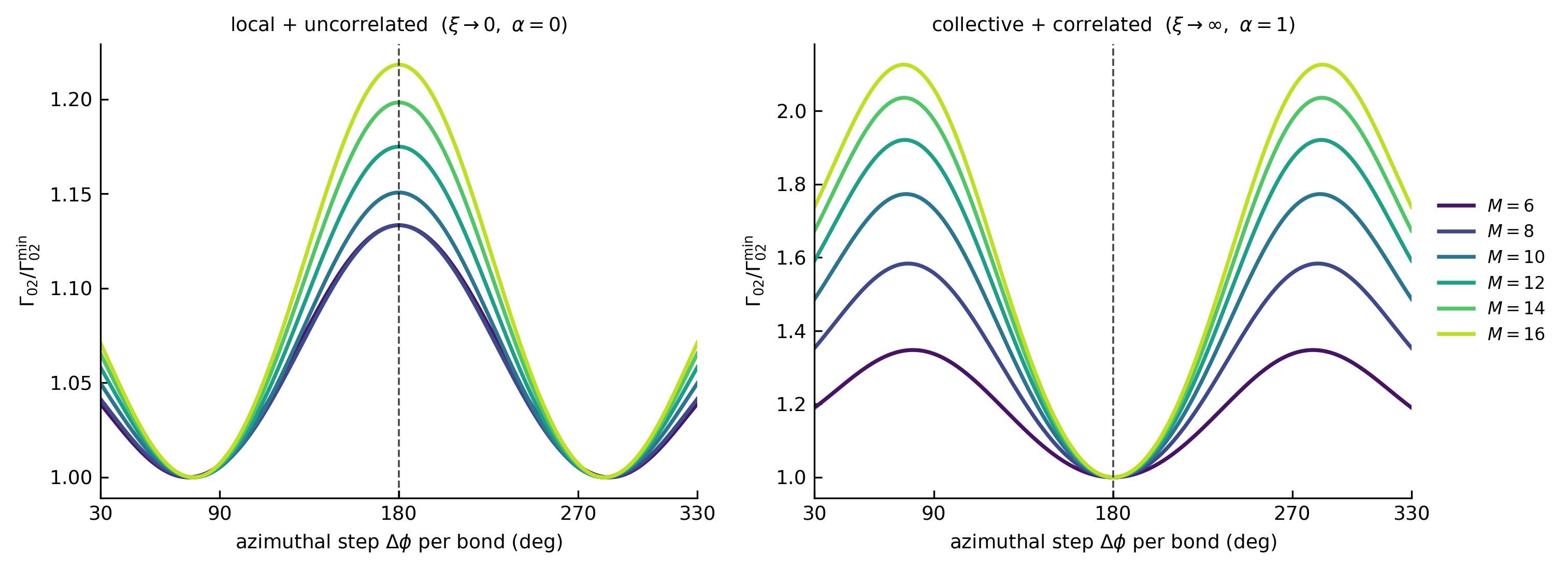}
    \caption{Normalized decoherence rate $\Gamma_{02}/\Gamma_{02}^{\mathrm{min}}$ as a function of azimuthal step $\Delta\phi$ for chain lengths $M = 6$--$16$ ($\lambda_0/t_0 = 0.3$) in (a) a local, uncorrelated bath ($\alpha = 0$, $\xi/\ell \to 0$) and (b) a collective, correlated bath ($\alpha = 1$, $\xi/\ell \to \infty$). For all chain lengths, the achiral geometry ($\Delta\phi = 180^\circ$; dashed line) is a maximum in (a) and a minimum in (b).}
    \label{fig:chain-length}
\end{figure*}
Similar argumentation may be used to analyze the population relaxation contribution to decoherence. In this case, the relevant quantity is the fluctuating bath-induced coupling between eigenstates $i$ and $l$:
\begin{equation}
    \delta \hat{V}_{il}(t) = \hat{\Lambda}_{il}(t)=\sum_{m,c}\braket{i|\hat{S}_m^c|l}\hat{B}_m^c(t).
\end{equation}
and the population-transfer rate constant from state $i$ into state $l$ is given by:\cite{kuhn_dynamics_2023}
\begin{equation}
    k_{i\to l} = \frac{2}{\hbar^2}\text{Re}\int_0^\infty dt e^{i\omega_{il}t}\langle\hat{\Lambda}_{il}(t)\hat{\Lambda}_{li}(0)\rangle.
\end{equation}
Then, the population-relaxation contribution to the decay of the $0$-$2$ coherence is 
\begin{equation}
        \Gamma_{02}^\text{rel} = \frac{1}{2}\left(\sum_{l\neq 0}k_{0\to l}+\sum_{l\neq 2}k_{2\to l}\right). 
\end{equation}
Again, noting that 
\begin{equation}
    \hat{\Lambda}_{il}=\sum_{m,c}\braket{i|\hat{S}_m^c|l} \hat{B}_{m}^c
\end{equation}
we obtain
\begin{equation}
    \langle \hat{\Lambda}_{il}(t)\hat{\Lambda}_{li}(0)\rangle = \sum_{m,m'}\sum_{c,c'}\braket{i|\hat{S}_m^c|l} \braket{i|\hat{S}_{m'}^{c'}|l}^* \langle\hat{B}_m^c(t)\hat{B}_{m'}^{c'}(0)\rangle
\end{equation}
so that our rate expression is 
\begin{equation}
    \begin{aligned}
        k_{i\to l} &= \frac{2}{\hbar^2} \sum_{m,m'}\sum_{c,c'}\braket{i|\hat{S}_m^c|l} \braket{i|\hat{S}_{m'}^{c'}|l}^* \int_0^\infty dt e^{i\omega_{il}t}\langle\hat{B}_m^c(t)\hat{B}_{m'}^{c'}(0)\rangle \\& = \frac{2}{\hbar^2}f_T(\omega_{il})\sum_{m,m'}\sum_{c,c'}\braket{i|\hat{S}_m^c|l} \braket{i|\hat{S}_{m'}^{c'}|l}^* J_{mm'}^{cc'}(|\omega_{il}|). 
    \end{aligned}
\end{equation}
where 
\begin{equation}
    f_T(\omega) = \begin{cases}
        n_B(\omega)+1 & \omega>0\\
        n_B(|\omega|) & \omega < 0 
    \end{cases}
\end{equation}
and 
\begin{equation}
    n_B(\omega) = \frac{1}{e^{\beta \hbar \omega}-1}.
\end{equation}
Substituting our spectral density, Eq. \eqref{spectral-density}, we obtain 
\begin{equation}\label{relaxation}
    \begin{aligned}
        k_{i\to l} & = \frac{2}{\hbar^2}|\omega_{il}|e^{-|\omega_{il}|/\omega_c}f_T(\omega_{il})\\
        &\quad \times \sum_{m,m'}\sum_{c,c'}\braket{i|\hat{S}_m^c|l} \braket{i|\hat{S}_{m'}^{c'}|l}^* \sqrt{A_cA_{c'}}K_{cc'}F_{mm'}. 
    \end{aligned}
\end{equation}
From this equation, we see two sources of geometry dependence on the decoherence rate. First, because the Hamiltonian 
$\hat{\mathcal{H}}_S$
is dependent on system geometry, as the geometry changes, the transition frequency $\omega_{il}$ between eigenstates $\ket{i}$ and $\ket{l}$ itself changes. As seen in Figure \ref{fig:mechanism}b for the $0\to 2$ transition, this results in a modulation of the prefactor in (\ref{relaxation}). Note, however, that this source of geometry dependence is independent of bath correlations. The double sum in (\ref{relaxation}), on the other hand, has the same structure as the double sum in (\ref{pure-dephasing}). When the bath is completely uncorrelated {($\alpha = 0$, $\xi/\ell \to 0$)} , the coupling terms are squared before being summed; when the bath is completely correlated {($\alpha = 1$, $\xi/\ell\to \infty$)}, the terms are summed before being squared, allowing for cancellation (see Figure \ref{fig:mechanism}c). 

The relative contributions of pure dephasing and population relaxation to the overall decoherence rate vary between the fully correlated and fully uncorrelated bath. In the case of the local, uncorrelated bath, pure dephasing and population relaxation both contribute substantially to the overall rate of decoherence. On the other hand, in the case of the collective, correlated bath, pure dephasing is the only contribution to decoherence. This is because in the limit $\xi \to \infty$ and $\alpha\to 1$, we may write 
\begin{equation}
   \begin{aligned}
        \hat{ \mathcal{H}}_{SB}
&=
\left(
\sqrt{A_t}\hat S_2+
\sqrt{A_\lambda}\hat S_3
\right)\otimes\hat B \\&= \left(
\sqrt{A_t}\hat S_2+
\sqrt{A_t} \frac{\lambda_0}{t_0}\hat S_3
\right)\otimes\hat B\propto \sqrt{A_t} \hat{\mathcal{H}}_S\otimes \hat{B}
   \end{aligned}
\end{equation}
for an appropriate bath operator $\hat{B}$ since $\eta = 1$ so that population relaxation must vanish identically. For other values of $\eta$, population relaxation may contribute to the overall rate of decoherence even in a collective, correlated bath although pure dephasing dominates in this regime. Nevertheless, it is clear that the overall trend in geometry-dependent decoherence is shaped by the relative contributions of pure dephasing and population relaxation. The geometry dependence of these two rates are, in turn, goverened by the correlation structure of the system-bath coupling. 

\subsection{Geometric control is robust to helix length}

Although the preceding analysis was done for the case $M=10$, the general trends hold across helix lengths. Figure \ref{fig:chain-length} shows the normalized decoherence rate $\Gamma_{02}/\Gamma_{02}^\text{min}$ as a function of azimuthal step $\Delta \phi$ for the local, uncorrelated bath and the collective, correlated bath for lengths $M = 6$--$16.$ We notice that across this range of lengths the same qualitative behavior is seen: $\Delta \phi= 180^\circ$ is a decoherence maximum for the local, uncorrelated bath and a minimum for the collective, correlated bath. Furthermore, the location of the chiral extrema is seen to stay roughly constant over this length range even as the number of helical turns changes by nearly a factor of three. Specifically, the chiral extremum (a minimum in the local, uncorrelated bath and a maximum in the collective, correlated bath) remains at $\Delta\phi \approx 74$--$80^\circ$. This indicates that the geometric response is determined by the local twist between adjacent sites rather than by the global shape of the helix, so that $\Delta\phi$ is a transferable design variable across helices of different lengths. We note, however, that the magnitude of the geometric response grows with $M$; our claim of robustness therefore concerns the sign of the response and the location of its extrema rather than its size.

\section{Conclusions}
In this work, we have demonstrated that the chiral geometry of a molecular system can serve as a continuous control parameter governing the rate of electronic decoherence of a prepared superposition of energy eigenstates. By studying a chiral tight-binding model, we showed that this geometry-dependence of decoherence rates is  controlled both by system properties and the nature of the bath and system-bath coupling. In particular, the magnitude of geometric control is modulated by the SOC strength of the system and is seen to be appreciable only in the intermediate- and strong-SOC regimes. On the other hand, the actual nature of geometry dependence is contingent on the correlation structure of the bath to which the system is coupled. In fact, different spatial and channel correlations were observed to give rise to inverted geometry dependence. 

These findings give rise to a number of practical consequences important to the design of quantum systems in room temperature condensed phase environments. First and foremost, it demonstrates that, in principle, chirality can serve as a tunable parameter that can control decoherence rates. Because this tunability of twist is experimentally realizable even among particles small enough to give rise to quantum effects, this work suggests a novel method of utilizing system geometry to modulate decoherence. 

Conversely, measuring how the decoherence rate varies with helical twist
could provide information about the correlation structure of the bath.
For the room-temperature parameters considered here, observing that the
prepared coherence decays most slowly at the achiral geometry would
suggest a long spatial correlation length and strong correlations
between fluctuations in the hopping and SOC channels. On the other hand, observing a chiral geometry decohere more slowly than the achiral geometry would suggest relatively short spatial and channel correlations. 

More broadly, this work extends the study of chirality–property relationships to the lifetime of electronic coherence. Our results suggest that the twist that best preserves coherence depends on how the environment couples to the helix. Measuring how this preferred twist changes across environments could help identify the correlations that govern decoherence and guide the design of chiral systems with longer-lived electronic coherence.
\section{Acknowledgements}

N.M. dedicates this work to the memory of his late grandfather, Ramchandra Babu. We thank members of the Kotov and Geva groups for helpful discussions. This work was supported by DOE grant DE-SC0026608. We acknowledge the use of Claude Opus 5.5 to assist in writing and debugging portions of the software scripts used in this study; all outputs and code logic were independently validated by the authors. 
\section{References}
\bibliography{references}

\end{document}